# Integrability and the variational formulation of non-conservative mechanical systems

## D. H. Delphenich[†]

Lindsborg, KS 67456

It is shown that one can obtain canonically-defined dynamical equations for non-conservative mechanical systems by starting with a first variation functional, instead of an action functional, and finding their zeroes. The kernel of the first variation functional, as an integral functional, is a 1-form on the manifold of kinematical states, which then represents the dynamical state of the system. If the 1-form is exact then the first variation functional is associated with the first variation of an action functional in the usual manner. The dynamical equations then follow from the vanishing of the dual of the Spencer operator that acts on the dynamical state. This operator, in turn, relates to the integrability of the kinematical states. The method is applied to the modeling of damped oscillators.

## 1  Introduction

One of the most, if not the most, fundamental principles of all physics is the least-action principle, which states that nature favors the state of least action, if one defines the terms "state" and "action" in the correct way. Applications of this principle can take the form of the path of least distance, the path of least time, the surface of least area, the deformation of an elastic body with the lowest deformation energy, and more general fields that minimize the action functional. The least-action principle then has the advantage of singling out an often unique, canonical, element from what usually amounts to an infinite-dimensional space of possibilities.

However, there are limits to the applicability of the least-action principle. For instance, in physical mechanics, one can only deal with conservative systems, which leaves out many common examples, such as mechanical systems that include friction or viscous drag, as well as open systems, in which energy is added or subtracted from "external" sources. Actually, one can regard friction and drag as a type of external energy source or sink in the sense that one is dealing with forms of energy that are beyond the scope of the model, except as essentially stochastic contributions.

---

[†] E-mail:  david_delphenich@yahoo.com

Another common physical milieu in which the least-action principle apparently breaks down – if only temporarily – is in quantum physics. One of the more popular approaches to the dynamics of quantum systems – i.e., systems of interacting atomic or subatomic particles – is the Feynman path integral, or more generally, the functional integral approach. With this technique, one basically suggests that the transition probability from a given incoming scattering state to a given outgoing one involves contributions from more than just the extremal path that connects them. This leads to the notion of a "loop" expansion, which puts the contribution from the extremal solution at the zero loop, or "tree" order of approximation and adds increasingly non-classical contributions from higher-loop diagrams in a manner that is analogous to the asymptotic expansions of diffraction theory in geometrical optics. However, one can often arrive at an "effective" action functional whose extremal solutions then represent the quantum-corrected classical extremals.

Much of the methodology of quantum field theory is by analogy with corresponding concepts in non-equilibrium thermodynamics, in which one addresses the excitations of the equilibrium – or ground – state from the standpoint of both small perturbations, which defines the linear theory, and large excitations, which defines the nonlinear theory, and is not always perturbative in character. For the instance, the methodology of spontaneous symmetry breaking and phase transitions comes out of this nonlinear theory of non-equilibrium thermodynamics.

One thing that becomes clear in the study of conservative mechanical systems is that invariably all that one is dealing with are essentially "accounting" principles. That is, whether or not some physical quantity that is associated with a mechanical system is conserved has a lot to do with the degree of completeness in one's statement of the various forms that the quantity can take in the system. For instance, when one first encounters elementary collisions in physics, one is told that there are two types of collisions: elastic ones, in which total energy is conserved and inelastic ones, in which it is not. One then learns that the first law of thermodynamics is that total energy is *always* conserved. The way that one resolves this contradiction is to understand that the total energy in the collision of hard spheres moving on frictionless surfaces included only the kinetic energy, but not, for instance, the binding energy that is absorbed if they stick together or the energy of deformation.

Hence, one is always dealing with the issue of the completeness of the system definition, and whether the system interacts with another unmodeled system. This unmodeled system can be external, in the sense of a heat bath or more astronomical systems, or internal, such as motion and energy at the atomic level when one is modeling a macroscopic process.

Now, one of the unavoidable realities of all logical systems, such as the laws of nature and the basic premises of any physical model, is Gödel's theorem that a logical system can be at best either logically consistent or logically complete, but not both. That is, if, as scientists always insist upon, the basic premises of science do not lead to any contradictions then there must be well-posed questions whose truth or falsehood cannot be resolved by that system. This suggests that any conservation law represents a level of approximation, just as symmetry and homogeneity are usually introduced into a model as an approximation for something more intractable.

Hence, it might be more prudent to deal with conservation laws in their "strong" form instead of their "weak" form. The terminology that we are using is that the weak statement of a conservation law is that the total amount of some physical quantity – e.g., mass, energy, momentum, angular momentum – remains constant in the course of the time evolution of the state of the system. The strong statement of that conservation law would be that the time derivative of that quantity is equal to some other specified quantity; one also refers to such laws as *balance laws* in continuum mechanics, thermodynamics, and statistical mechanics. For instance, Newton's first law of motion becomes the weak form of the law of conservation of momentum and his second law, when expressed in the form $\sum \mathbf{F} = d\mathbf{p}/dt$, gives the strong form of that law, or balance law.

What we are proposing to do in this study is to show that there is a physically useful expansion of scope of the least-action principle that exhibits the extremal motion of classical mechanics as being like the weak form of a conservation law when there is a strong form that includes it as a limiting approximation. The key to presenting a statement of that more general principle is to formulate the conventional calculus of variations in the language of jets and then show how one can use the integrability of sections of these fibered manifolds to deduce equations that subsume the Euler-Lagrange equations that one derives by starting with a Lagrangian function $\mathcal{L}$, but without actually having to start with a Lagrangian. Indeed, one basically replaces the exact 1-form $d\mathcal{L}$ with a more general 1-form $\phi$, as we shall see.

In section 2, we summarize the relevant facts from the formulation of the calculus of variations in terms of jets that we will generalize. In section 3, we will make the generalization of these constructions that will enable us to address the mechanics of non-conservative systems. In section 4, we briefly describe the relationship between the dynamical equations that we defined in section 3 and the integrability of sections of the projection of the jet manifold onto the parameter space. In section 5, we give examples of how to apply the general method to the modeling of damped one-dimensional oscillators with both linear and nonlinear restoring and damping forces. Finally, in section 6, we summarize the key points of the method and propose the immediate directions for further development.

Many of the ideas that are presented below have been discussed in greater mathematical and physical detail in Delphenich [1]. One will find that the method that is proposed here of deriving dynamical equations from the dual of the Spencer operator is a generalization of the method that was proposed by Pommaret [2], which still began with the definition of an action functional.

## 2 The calculus of variations in the language of jets [1]

The *1-jet* of a differentiable function $f$: $M \to N$ at a point $u \in M$ is defined to be the equivalence class of all differentiable functions that are defined in some neighborhood of

---

*u* and have the same value as a function at *u* as *f* – i.e., *f*(*u*) – and the same value of their derivative at *u* as $df|_u$. We then denote this equivalence class by $j^1 f_u$. Note that since it is not necessary for the functions to be globally defined it is still possible to speak of 1-jets when global functions do not exist, such as 1-jets of sections of principle fiber bundles when the bundle in question is not trivial. For instance, one can define 1-jets of local frame fields on manifolds that are not parallelizable.

The disjoint union $J^1(M, N)$ of all 1-jets at all of the points of *M* projects onto *M*, *N*, and $M \times N$ by way of the maps $\pi_M$, $\pi_N$, and $\pi_{1,0}$ that take $j^1 f_u$ to $u \in M$, $f(u) \in N$, and (*u*, *f*(*u*)) $\in M \times N$, respectively; one gives $J^1(M, N)$ the topology that is induced by the projection in the last case. Furthermore, one can give it a differential structure by starting with a chart (*U*, $u^a$) about any $u \in M$ and a chart (*V*, $x^i$) about any $x \in N$ and defining a chart on a subset of $\pi_{1,0}(U \times V)$ by the coordinates $(u^a, x^i, x^i_a) \in \mathbb{R}^m \times \mathbb{R}^n \times \mathbb{R}^{mn}$. We shall then refer to $J^1(M, N)$ with this differential structure as the *manifold of 1-jets of differentiable maps from M to N*.

The projections $\pi_M : J^1(M, N) \to M$ and $\pi_N : J^1(M, N) \to N$ do not define fibrations, but only fibered manifolds; that is, these projections are onto and so are their differential maps. However, the projection $\pi_{1,0} : J^1(M, N) \to M \times N$ is a fibration, and the typical fiber is an affine space that is modeled on $\mathrm{Hom}(\mathbb{R}^m, \mathbb{R}^n)$, that is, the vector space of linear maps from $\mathbb{R}^m$ to $\mathbb{R}^n$, which is linearly isomorphic to $\mathbb{R}^{mn}$. One can then think of the elements of the resulting affine bundle as representing the differential parts of the 1-jets, since the fiber coordinates are the $x^i_a$ in this case.

A *section* of the projection $\pi_M : J^1(M, N) \to M$ is then a map $\psi : M \to J^1(M, N)$ that takes each $u \in M$ to some 1-jet $\psi(u)$ in the fiber over *u*. Hence, one must have that the composition $\pi_M \cdot \psi$ is the identity map on *M*. The local form for the image $\psi(U)$ of a local section $\psi : U \to J^1(M, N)$ when $\psi(U)$ is contained in a chart on $J^1(M, N)$ is:

$$\psi(u) = (u^a, x^i(u), x^i_a(u)). \tag{2.1}$$

It is essential in what follows to clearly distinguish this general section from the more particular case of a section of $\pi_M$ that is the *1-jet prolongation* $j^1 f: M \to J^1(M, N)$, $u \mapsto j^1 f_u$, of a differentiable map from *M* to *N*. It will take the local form:

$$j^1 f(u) = (u^a, x^i(u), x^i_{,a}(u)), \tag{2.2}$$

in which the comma denotes partial differentiation with respect to the independent variable $u^a$. If $\psi$ takes the form of $j^1 f$ for some differentiable map *f* then one says that $\psi$ is *integrable*.

We shall return to this issue in a later section in which we account for our extension of the least-action principle. First, however, we shall summarize the usual calculus of variations, at least for first order Lagrangians, as it is expressed in the language of 1-jets.

For the sake of convenience, we assume that *M*, which we now call *K*, is a compact connected subset of $\mathbb{R}^m$, which then plays the role of a parameter space for whatever object the embedding of $K$ in $N$ would represent. For instance, in point mechanics, one might use $K = [a, b] \subset \mathbb{R}$, which represents a finite proper time parameter interval [2]. The compactum $K$ might or might not have a non-vacuous boundary $\partial K$.

In order to define an action functional on the differentiable maps of $K$ into M, one first chooses a differentiable function $\mathcal{L}: J^1(K, M) \to \mathbb{R}$ that one calls the *Lagrangian density* for the action functional. Hence, it will take the local form $\mathcal{L}(u^a, x^i, x^i_a)$.

If $V \in \Lambda^m(K)$ is a volume element on $K$ then one can lift it to an $m$-form $\hat{V} \in \Lambda^m(J^1(K, M))$ by means of the projection $\pi_M$. The scalar multiple $\mathcal{L}\hat{V}$ is another $m$-form on $J^1(K, M)$, and when one chooses a section $\psi: K \to J^1(K, M)$ one can pull $\mathcal{L}\hat{V}$ down to the $m$-form $\mathcal{L}V$ on $K$. If this seems a bit roundabout, keep in mind that we will need to exterior differentiate the $m$-form $\mathcal{L}\hat{V}$, and if we did that on $K$ then we would only get zero, since all $m+1$-forms on an $m$-dimensional manifold are null.

We then define the *action functional* on differentiable maps $f: K \to N$ to be the association of $f$ with the number:

$$S[f] = \int_K \mathcal{L}(j^1 f)\hat{V} = \int_K \mathcal{L}(u^a, x^i(u), x^i_{,a}(u))\, du^1 \wedge \cdots \wedge du^m \ . \tag{2.3}$$

Although it is not always analytically rigorous, or at least computationally useful, to regard the set $C^1(K, N)$ as an infinite-dimensional differentiable manifold and the action functional as a differentiable function on that manifold, it is certainly heuristically useful to imagine that what the calculus of variations really represents is the "calculus of infinity variables," or, at least, the theory of critical points of differentiable functions on infinite-dimensional differentiable manifolds.

In particular, if $f: K \to N$ is a "point" of this "manifold" then a "tangent vector" at $f$ takes the form of a vector field $\delta x$ on $f(K)$. One also refers to this vector field as the *first variation* of $f$, since it will be the infinitesimal generator of a one-parameter family of diffeomorphisms of $f(K)$ into $M$. It will then have the local form:

$$\delta x = \delta x^i \frac{\partial}{\partial x^i} \ . \tag{2.4}$$

Hence, we are not assuming that the parameter space $K$ is being deformed as a result of $\delta x$.

The first variation $\delta x$ of $f$ then has a prolongation to a vector field $\delta^1 x$ on $J^1(K, M)$, which has the local form:

$$\delta^1 x = \delta x^i \frac{\partial}{\partial x^i} + \frac{\partial(\delta x^i)}{\partial u^a} \frac{\partial}{\partial x^i_a} \ . \tag{2.5}$$

---

We can then define the *first variation of the action functional* at $f$ to be a "1-form" on our "manifold" $C^1(K, N)$ that takes the "tangent vector" $\delta^1 x$ to the number:

$$\delta S|_f[\delta^1 x] = \int_K L_{\delta^1 x}(\mathcal{L}\hat{V}) = \int_M i_{\delta^1 x}(d\mathcal{L}\wedge\hat{V}) = \int_K d\mathcal{L}(\delta^1 x)\hat{V}\,, \tag{2.6}$$

in which $L_{\mathbf{X}} = i_{\mathbf{X}}d + di_{\mathbf{X}}$ is the Lie derivative operator with respect to the vector field $\mathbf{X}$, as it acts on differential forms.

Since $d\mathcal{L}$ will take the local form:

$$d\mathcal{L} = \frac{\partial\mathcal{L}}{\partial x^i}dx^i + \frac{\partial\mathcal{L}}{\partial x_a^i}dx_a^i \equiv F_i\,dx^i + \Pi_i^a dx_a^i\,, \tag{2.7}$$

if $\delta^1 x$ takes the local form that was given in (2.5) then we can say that:

$$\delta S|_f[\delta^1 x] = \int_K \left( F_i\delta x^i + \Pi_i^a\frac{\partial(\delta x^i)}{\partial u^a}\right) V\,. \tag{2.8}$$

Note that it is not necessary to include the contribution to $d\mathcal{L}$ from $\partial\mathcal{L}/\partial u^a\,du^a$ is because it will be annulled by exterior multiplication with $\hat{V}$.

By the product rule for differentiation, which is usually referred to as an integration by parts, (2.8) can be given the form:

$$\delta S|_f[\delta^1 x] = \int_K \frac{\delta\mathcal{L}}{\delta x}(\delta x)V - \int_{\partial K}\Theta(\delta x)\,, \tag{2.9}$$

in which we have introduced the *variational derivative* 1-form $\delta\mathcal{L}/\delta x$ and the *transversality* $m-1$-form $\Theta$, which are defined by:

$$\frac{\delta\mathcal{L}}{\delta x} = \left( F_i - \Pi_{i,a}^a\right)dx^i\,, \qquad \Theta = \Pi_i^a dx^i\otimes\#\partial_a\,. \tag{2.10}$$

The notation $\#\partial_a$ refers to the $m-1$-form on $K$ that is Poincaré dual to the vector field $\partial_a$, namely:

$$\#\partial_a = i_{\partial/\partial u^a}V\,. \tag{2.11}$$

The classical problems in the calculus of variations take the form of finding extrema of the action functional, which are then the "points" $f$ at which the first variation "1-form" has a zero, at least when restricted to "tangent vectors" $\delta x$ that satisfy either the *fixed-boundary condition* that $\delta x = 0$ when restricted to $\partial K$, or the *transversality condition* that $\Theta(\delta x) = 0$ when restricted to $\partial K$, which then allows one to pose *free-boundary* problems, as well. In either case, the boundary integral in (2.9) vanishes, and one finds that $f$ is an extremum for $S[.]$ – i.e., a critical point – iff $\delta S|_f[\delta x] = 0$ for all $\delta x$ that satisfy the specified boundary conditions iff the *Euler-Lagrange* equations are satisfied:

$$\frac{\delta \mathcal{L}}{\delta x} = 0, \tag{2.12}$$

which can also be put into the form:

$$F_i = \Pi^a_{i,a}, \tag{2.13}$$

which take the form of Newton's second law when $K = [a, b]$, so $\Pi^a_i$ takes the form of the components $p_i = \partial \mathcal{L} / \partial \dot{x}^i$ of the momentum 1-form. In the case of $\dim(K) > 1$, these equations generalize either the fundamental equations of continuum statics or dynamics, depending upon the way that one interprets $K$.

### 3  Zeroes of more general first variation functionals

As we said in the last section, an action functional $S[.]$ can be regarded, at least formally, as a differentiable function on an infinite-dimensional "manifold" $C^1(K, N)$ of states. By differentiation, one obtains a "1-form" $\delta S[.]$ on that manifold, which then takes a "tangent vector" $\delta \psi$ at a kinematical state $\psi$ – i.e., a variation of that state – to a number $\delta S[\delta \psi]$ that represents the directional derivative of $S$ in the direction $\delta \psi$. An extremal state then becomes a critical point of $S$, i.e., a zero of the 1-form $\delta S$.

A physically useful generalization of this is to define the first variation functional in terms of a 1-form $\phi \in J^1(K, M)$ that is not exact, and is therefore not derived from a Lagrangian function, in place of the $d\mathcal{L}$ that we used in (2.6). The first variation functional is essentially an infinite-dimensional analogue of a non-conservative force 1-form, which therefore does not admit a potential function, and, in fact, that is precisely how we will introduce such forces into a variational formulation of non-conservative motion.

The main question to resolve is how to obtain equations of motion that are canonical in some sense when we are no longer interpreting the zeroes of $\delta S|_f$ as extrema of an action functional on $f$. As we shall see, the solution to this dilemma is to be found in the integrability of the 1-form $\phi$, in a different sense than its integrability under the exterior derivative.

The way that we define our generalization of $\delta S$ is as follows: let $\phi$ be a 1-form on $J^1(K, N)$ that is vertical for the projection $J^1(K, N) \to K$. Hence, it can be represented locally in the form:

$$\phi = F_i dx^i + \Pi^a_i dx^i_a, \tag{3.1}$$

as in (2.7).

If $\hat{V}$ is the lift of the volume element $V \in \Lambda^m(K)$ to $J^1(K, N)$ then one can form the $m+1$-form $\Omega = \phi \wedge \hat{V}$ on $J^1(K, N)$. Now, let $\psi: K \to J^1(K, N)$ be a section of $J^1(K, N) \to K$ and let $\delta\psi$ be a vector field on $J^1(K, N)$ that is tangent to $\psi$. Hence, it will have the local form:

$$\delta\psi = \delta x^i \frac{\partial}{\partial x^i} + \delta x_a^i \frac{\partial}{\partial x_a^i}. \tag{3.2}$$

One can then take the interior product $i_{\delta\psi}\Omega$ and obtain an $m$-form on $\psi$ that will have the local form:

$$i_{\delta\psi}\Omega = (F_i \delta x^i + \Pi_i^a \delta x_a^i)\hat{V}. \tag{3.3}$$

One can then pull the $m$-form $i_{\delta\psi}\Omega$ down to an $m$-form $\psi^*(i_{\delta\psi}\Omega)$ by way of $\psi$. The only essential difference between its local form and the one in (3.3) is that all of the component functions for $\phi$ and $\delta\psi$ will be functions of $\psi(u)$.

This $m$-form on $K$ can be integrated over $K$ to obtain a number:

$$\Sigma|_\psi[\delta\psi] = \int_K i_{\delta\psi}\Omega, \tag{3.4}$$

and one sees that for a given $\psi$ the association of $\delta\psi$ with $\Sigma|_\psi[\delta\psi]$ is linear. Hence, $\Sigma|_\psi$ is essentially a "1-form" on the "tangent vectors" $\delta\psi$ to the "manifold" of all sections $\psi$ at the "point" $\psi$, and we call it the *first variation functional*.

One immediately finds that one can duplicate the steps that led to the Euler-Lagrange equations that one obtained from a Lagrangian $\mathcal{L}$ without the necessity of having to introduce one. The key step, viz., the "integration by parts", depends only upon the assumption that the variation $\delta\psi$ is a prolongation $\delta^1 x$ of a vertical vector field $\delta x$ on $f(K)$. However, if one chooses a variation $\delta\psi$ that is not integrable then such a variation would be expressible in the form:

$$\delta\psi = \delta x^i \frac{\partial}{\partial x^i} + \left( \delta x_{,a}^i - D\delta x_a^i \right) \frac{\partial}{\partial x_a^i}, \tag{3.5}$$

in which we have defined:

$$D\delta x_a^i = \delta x_{,a}^i - \delta x_a^i. \tag{3.6}$$

We now get:

$$i_{\delta^1 x}\Omega = \left( F_i \delta x^i + \Pi_i^a \frac{\partial(\delta x^i)}{\partial u^a} - \Pi_i^a D\delta x_a^i \right)\hat{V}. \tag{3.7}$$

By substituting this into (3.4), one obtains:

$$\Sigma|_\psi[\delta^1 x] = \int_K \left( F_i \delta x^i + \Pi_i^a \frac{\partial (\delta x^i)}{\partial u^u} - \Pi_i^a D \delta x_a^i \right) \hat{V} \,. \tag{3.8}$$

whose right-hand side is essentially the same as in (2.8).

By an integration by parts this then takes the form:

$$\Sigma|_\psi[\delta^1 x] = \int_K \left[ \left( F_i - \frac{\partial \Pi_i^a}{\partial u^a} \right) \delta x^i - \Pi_i^a D \delta x_a^i \right] \hat{V} + \int_{\partial K} (\Pi_i^a \delta x^i) \# \partial_a \,, \tag{3.9}$$

which we then write as:

$$\Sigma|_\psi[\ \delta^1 x] = \int_K [D^* F(\delta x) - \Pi_i^a (D \delta x_a^i)] \hat{V} + \int_{\partial K} \Theta(\delta x) \,, \tag{3.10}$$

in which we have defined the 1-form $D^* F$ on $J^1(K, N)$ by:

$$D^* F = \left( F_i - \frac{\partial \Pi_i^a}{\partial u^a} \right) dx^i \,, \tag{3.11}$$

and $\Theta$ is the transversality 1-form that we defined in (2.10).

Hence, one can also regard $\Sigma|_\psi$ as a linear functional on the subspace of the "tangent space" at $\psi$ that is spanned by vector fields on $\psi(K)$ of the form (3.2).

One sees that if one looks for the zeroes of $\Sigma|_\psi$ under the usual variational restrictions that $\delta \psi$ be the prolongation of a variation $\delta x$ – so $D \delta x_a^i = 0$ – and that $\delta x$ make the boundary contribution vanish, namely, the vanishing of $\delta x$ on $\partial K$ or the transversality condition $\Theta(\delta x) = 0$, one finds that $\Sigma|_\psi[\delta x] = 0$ for all such $\delta x$ iff:

$$D^* F = 0, \tag{3.12}$$

which takes the local form (2.13).

As we shall explain in the next section, in which we discuss the origin of the $D^*$ operator in the integrability of sections of $J^1(K, N)$, if we define $D^* \Pi = 0$, identically, then we can extend (3.12) to the equation:

$$D^* \phi = 0 \,, \tag{3.13}$$

which we will regard as the definitive one.

In order to relate this construction to the classical variational constructions in section 2, one need only set $\phi = d\mathcal{L}$, where $\mathcal{L}$ is a Lagrangian density function on $J^1(K, N)$. For such a $\phi$, one then has:

$$D^* F = \frac{\delta \mathcal{L}}{\delta x} \,, \tag{3.14}$$

and (3.12) is, indeed, a generalization of the Euler-Lagrange equations defined by $\mathcal{L}$.

Furthermore, when $\phi$ is not exact, it is often possible, at least locally, to uniquely decompose $\phi$ into a sum $d\mathcal{L} + \phi_a$, where $\mathcal{L}$ is some $C^1$ function on $J^1(K, N)$ and $\phi_a$ is not exact. (See the section on the Poincaré lemma in Bott and Tu [**7**])

The way that one does this in a coordinate chart on $J^1(K, N)$ is by means of locally-defined *cochain homotopy operator H*: $\Lambda^1(J^1) \to \Lambda^0(J^1)$, where:

$$\mathcal{L}(u^a, x^i, \; x_a^i) = H\phi(u^a, x^i, \; x_a^i) = \int_0^1 (i_{\mathbf{R}}\phi)(s\mathbf{R})ds \; , \tag{3.15}$$

in which:

$$\mathbf{R} = u^a \frac{\partial}{\partial u^a} + x^i \frac{\partial}{\partial x^i} + x_a^i \frac{\partial}{\partial x_a^i} \tag{3.16}$$

is the "radius" or "position" vector field (centered at the origin) that is defined on $\mathbb{R}^m \times \mathbb{R}^n \times \mathbb{R}^{mn}$ for the chosen coordinate chart.

The decomposition of $\phi$ into an exact part $d\mathcal{L}$ and an "anti-exact" part $\phi_a$, which is annulled by $H$, follows from the basic property of $H$ that it define a cochain contraction:

$$I = dH + Hd, \tag{3.17}$$

which makes:

$$\phi = dH\phi + Hd\phi = d\mathcal{L} + \phi_a \; . \tag{3.18}$$

From the linearity of $D^*$, one then has:

$$D^*\phi = D^*(d\mathcal{L}) + D^*\phi_\alpha , \tag{3.19}$$

and equation (3.12) then says that:

$$D^*(d\mathcal{L}) = -D^*\phi_\alpha , \tag{3.20}$$

and when we take (3.14) into account, we obtain:

$$\frac{\delta\mathcal{L}}{\delta x} = -D^*\phi_\alpha , \tag{3.21}$$

which shows what happens to the Euler-Lagrange equations for a non-conservative system. The right-hand side of (3.21) then represents the contributions of external and non-conservative forces to the dynamical equations.

### 4 The integrability of sections of $J^1(K, N)$

As we pointed out above in section 2, not all sections $\psi\colon K \to J^1(K, M)$ represent the 1-jet prolongations of differentiable maps $f\colon K \to M$; i.e., not all of them are *integrable*. The condition for integrability can be expressed locally as a system of partial differential equations for the functions $x^i(u^a)$. If $\psi(u) = (u^a, x^i(u),\ x_a^i(u))$ then they are:

$$\frac{\partial x^i}{\partial u^a} = x_a^i. \tag{4.1}$$

One can either rephrase this in terms of a 1-form $\theta$ on $J^1(K, M)$ that takes its values in $T(M)$ and is called the *contact form* for $J^1(K, M)$, with the local representation:

$$\theta = \left(dx^i - x_a^i du^a\right)\frac{\partial}{\partial x^i}, \tag{4.2}$$

or in terms of the *Spencer operator* [3], which takes the form $D\colon J^1(K, M) \to T^*(K) \otimes J^0(K, M)$, $\psi \mapsto D\psi = j^1(\pi_{1,0}\psi) - \psi$, which locally looks like:

$$D\psi(u^a, x^i(u),\ x_a^i(u)) = (u^a, Dx^i(u)), \tag{4.3}$$

in which:

$$Dx^i = \left(\frac{\partial x^i}{\partial u^a} - x_a^i\right)du^a. \tag{4.4}$$

Here, we must point out that the Spencer operator really acts on sections of the projections involved in its definition, not the manifolds themselves, since differentiation is not defined, otherwise. We are also defining the manifold $J^0(K, N)$ to be $K \times N$, which then projects on $K$ and $N$ in the usual way, and a section of $J^0(K, N) \to K$ is then just a differentiable map from $K$ into $N$.

From (4.1), a section $\psi$ is then integrable iff either $\psi^*\theta = 0$ or $D\psi = 0$. The former condition is most convenient when one is dealing with systems of partial differential equations as exterior differential systems on $J^1(K, N)$, whereas the latter is more useful when on considers them in terms of fibered submanifolds of $J^1(K, N)$. Clearly, the two approaches are not independent of each other since the fibered submanifold in the latter case could very well be an integral submanifold of the exterior differential system that is defined in the former.

In order to account for the operator $D^*$ that we introduced in the last section we first have to extend to the next order of jets of maps from $K$ to $N$. The extension is straightforward: the *2-jet* $j^2 f_u$ of $f: K \to N$ at $u \in K$ is the equivalence class of all $C^2$ maps from some neighborhood of $u$ into $N$ that have the same values at $u$ as functions, along with their first and second differential maps. One then defines the manifold $J^2(K, N)$ in a manner that is analogous to the previous definition of $J^1(K, N)$, and it is fibered over $K$ and $N$, but not as a bundle, as before, although now its projection $\pi_{2,1}: J^2(K, N) \to J^1(K, N)$ defines an affine bundle whose fibers are modeled on the vector space $S^2(\mathbb{R}^m) \otimes \mathbb{R}^n$ of symmetric covariant tensors over $\mathbb{R}^m$ with values in $\mathbb{R}^n$, which serves as the space of second partial derivative components. A local coordinate chart on $J^2(K, N)$ then takes the form of $(u^a, x^i, x_a^i, x_{ab}^i)$.

---

[3] Donald Spencer first introduced it in his work [**8**] on the deformations of structures on manifolds that are defined by pseudogroups of transformations, and later applied it to the integrability of over-determined systems of linear differential equations [**9**]. This was extended to the nonlinear case by Hubert Goldschmidt in [**10**].

Similarly, we extend the Spencer operator to $D: J^2(K, N) \to T^*(K) \otimes J^1(K, N)$, $\psi \mapsto j^1(\pi_{2,1} \psi) - \psi$, which takes the local form:

$$D\psi = (u^a, Dx^i(u), Dx_a^i(u)),\qquad(4.5)$$

with $Dx^i(u)$ defined as before, and:

$$Dx_a^i = \left(\frac{\partial x_b^i}{\partial u^a} - x_{ba}^i\right) du^a.\qquad(4.6)$$

We now have to consider the dual of the Spencer operator, which will then be a map $D^*: T(K) \otimes T^*(J^1(K, N)) \to T^*(J^2(K, N))$, $\phi \mapsto D^*\phi$, that one obtains by pulling back a 1-form $\phi$ on $T^*(K) \otimes J^1(K, N)$ over $D\psi(K)$ to a 1-form $D^*\phi$ on $J^2(K, N)$ over $\psi(K)$.

Hence, if $\delta\psi = \delta x^i \partial / \partial x^i + \delta x_a^i \partial / \partial x_a^i$ is a vertical vector field on $J^1(K, N)$, which also defines a vector field on $J^2(K, N)$ with the same local form, then we should have:

$$D^*\phi|_\psi(\delta\psi) = \phi|_{D\psi}(\delta\psi);\qquad(4.7)$$

however, we shall see that this is true only up to a divergence.

If we evaluate the right-hand side of (4.7) locally in the general case of $\delta\psi = \delta x - D\delta\psi$ then we get:

$$
\begin{aligned}
\phi|_{D\psi}(\delta\psi) &= F_i \delta x^i + \Pi_i^a \frac{\partial(\delta x^i)}{\partial u^a} - \Pi_i^a D\delta x_a^i \\
&= \left(F_i - \frac{\partial \Pi_a^i}{\partial u^a}\right)\delta x^i + \frac{\partial(\Pi_a^i \delta x^i)}{\partial u^a} - \Pi_a^i D\delta x_a^i.
\end{aligned}\qquad(4.8)
$$

If the left-hand side of (4.7) takes the form:

$$D^*\phi|_\psi(\delta\psi) = D^* F_i \ \delta x^i + (D^* \Pi_i^a)\delta x_a^i\qquad(4.9)$$

then we see that if $\delta\psi$ is integrable then the two sides of (4.7) are consistent, up to a divergence, if we set:

$$D{*}F_i = F_i - \Pi_{i,a}^a, \qquad D^*\Pi_a^i = 0 .\qquad(4.10)$$

The divergence gives an exact form when we multiply both sides of (4.7) by $V$. Hence, by Stokes's theorem, it will contribute only a boundary term to the first variation 1-form under integration over $K$. Since this boundary term is the transversality $m-1$-form that we defined before, this contribution will vanish under the assumption of a fixed or transversal boundary variation.

We have thus accounted for the appearance of $D{*}F_i$ in the equations for the zeroes of the first variation functional $\delta\mathbb{S}[.]$. However, it must be observed that, unlike the exact case, in which it was sufficient to deal with 1-jets of maps, we have had to extend to 2-jets. This relates to the fact that the order of jets in the kinematical state of the system is equal to the order of the dynamical equations, which is one less than the order of the dynamical state. For more details on this, see Delphenich [1].

We now illustrate the results with some elementary physical examples.

## 5 Mechanical examples

Let us start with the forced, damped, one-dimensional harmonic oscillator, for the sake of specificity. Customarily, the basic dynamical principle is given by Newton's second law:

$$m\ddot{x} = \sum F = -kx - b\dot{x} + f(\tau),\qquad(5.1)$$

in which $k$ is the spring constant, $b$ is the damping constant, and $f(\tau)$ is the forcing function.

Now, the term on the left-hand side, when multiplied by the 1-form $d\dot{x}$, gives an exact 1-form:

$$m\dot{x}\,d\dot{x} = d(KE) = d\left(\tfrac{1}{2}m\dot{x}^2\right).\qquad(5.2)$$

When the terms in the right-hand side of (5.1) are multiplied by $dx$, only the first one is exact:

$$-kx\,dx = -dU = -d\left(\tfrac{1}{2}kx^2\right).\qquad(5.3)$$

Hence, we can define the exact part of $\phi$ by the undamped, unforced part of the oscillator:

$$d\mathcal{L} = d(-U + KE) = -kx\,dx + m\dot{x}\,d\dot{x} .\qquad(5.4)$$

The remaining part of $\phi$ is then:

$$\phi_o = (-\,b\dot{x} + f(\tau))\,dx = F_o\,dx.$$ (5.5)

We see immediately that:

$$d\phi_o = b\,dx \wedge d\dot{x} + \dot{f}\,d\tau \wedge dx\,.$$ (5.5)

Hence, $\phi_o$ is not closed, so it cannot be exact.

Since $\phi_o$ is independent of momentum, we then see that:

$$D^*\phi_o = (D^*F_o)\,dx = \phi_o\,.$$ (5.6)

Thus, the dynamical equations (3.21) that follow from our extended dynamical principle $D^*\phi = 0$ take the form:

$$\frac{\delta \mathcal{L}}{\delta x} = \phi_o\,,$$ (5.7)

which gives (5.1) upon substitution.

The restriction to a damped linear oscillator is not necessary in the foregoing, and one can easily extend to damped nonlinear oscillators of the Duffing and Van der Pol type (see Thompson and Stewart [11] or Guckenheimer and Holmes [12]). The best way to summarize the various possibilities it to point out that they are all based in giving a specific functional form to the components $F_i = F_i(u^a, x^i, x^i_a)$, which becomes $F = F(\tau, x, \dot{x}^i)$ when $K = [a, b]$. This basically amounts to defining a *mechanical constitutive law* that associates a dynamical state $\phi$ to the kinematical state $\psi$. One can also generalize the functional form of the components $\Pi^i_a = \Pi^i_a(u^a, x^i, x^i_a)$, which is essential in the case of continuum mechanics, but in the case of point mechanics, most commonly, one sets the momentum components $p_i$ equal to $m\dot{x}_i$, where the covelocity $\dot{x}_i$ is obtained from the velocity $\dot{x}^i$ by way of a spacetime metric; viz., $\dot{x}_i = g_{ij}\dot{x}^j$.

For instance, most of the traditional constitutive laws for one-dimensional motion take the form $F = f(\tau) + dU(x) + F_d(\dot{x})$, in which $f(\tau)$ is an external forcing term, $U(x)$ is a potential function, and $F_d(\dot{x})$ represents the damping force. Note that in dimension one all 1-forms are closed, hence, locally exact, so there is no inexact contribution to $F(x)$.

For the Duffing oscillator, one retains the linear damping, but replaces the linear restoring force with a force that is cubic in the displacement from equilibrium, so one has:

$$U(x) = \tfrac{1}{4}ax^4\,, \qquad F_d(\dot{x}) = b\dot{x}\,.$$ (5.8)

One can then think of this oscillator as a linearly damped anharmonic oscillator.

For the Rayleigh-Van der Pol oscillator, one retains the linear restoring force and replaces the linear damping force with a force that depends upon both $x$ and $\dot{x}$:

$$U(x) = \tfrac{1}{2}kx^2\,, \qquad F_d(x, \dot{x}) = b(x^2)\dot{x}\,.$$ (5.9)

This can be interpreted by saying that the damping coefficient $b$ varies symmetrically with the distance from the equilibrium position.

The situation in continuum mechanics is complicated by not only the fact that $\dim(K)$ > 1, which means replacing total derivatives with respect to $\tau$ with partial derivatives with respect to $u^a$, but also the fact that the kinematical state needs more explanation. We shall address these issues in a subsequent study, but, for now, we just point out that examples of how to apply the method to continuum mechanics include deformations of viscoelastic media and the Navier-Stokes equation in hydrodynamics.

## 6 Discussion

To summarize:

*i*. One can still arrive at canonically defined dynamical equations for non-conservative physical systems by variational methods if one starts, not with an action functional $S[.]$ that is based on a Lagrangian density, but with a first variation functional $\delta S[.]$ that is based on a 1-form $\phi$ that represents the dynamical state of the system on a jet manifold $J^1(K, N)$ that represents the kinematical state space of the system.

*ii*. The reduction to an action functional comes from using an exact dynamical 1-form $d\mathcal{L}$ for $\phi$.

*iii*. The specification of a particular functional form for the components of $\phi$ in terms of the kinematical state variables represents a choice of mechanical constitutive law for the system.

*iv*. The dynamical equations, $D^*\phi = 0$, which generalize the Euler-Lagrange equations, follow from the vanishing of the dual of the Spencer operator $D^*$ that acts on the dynamical state. This operator is rooted in the integrability of kinematical states.

*v*. When $\phi$ is decomposed into the sum of an exact 1-form $d\mathcal{L}$ and an inexact one $\phi_o$, the resulting form of the dynamical equations is $\delta\mathcal{L}/\delta x = -D^*\phi_o$, which applies to many of the common physical examples of dynamical equations for non-conservative systems, such as damped oscillators.

In addition to extending the physical examples to continuum mechanics, one can also extend the basic formalism to field theories in a straightforward way. One mostly has to specialize the manifold $N$ to a fiber bundle $N \to K$ over $K$ and specialize the differentiable maps from $K$ to $N$ to differentiable sections of that fibration. The jet manifold $J^1(K, N)$ then represents 1-jets of local $C^1$ sections of that bundle.

In the context of field theories, one must naturally address the issue of what happens to Noether's theorem. When one is dealing with non-conservative systems, it is reasonable to expect that the conserved currents associated with symmetries of the action functional give way to non-conserved currents that are associated with symmetries of the first variation functional. One also expects that when $\phi$ is decomposed into an exact form and an inexact one, the conservation law for the current takes the form $\delta\mathbf{J} = -$ (something obtained from the inexact part). These issues will also be dealt with in a subsequent study.

Another issue that must be addressed eventually is the relationship of the integrability of the first variation functional to the loop expansion of the effective actions in quantum field theory, which suggests a recursive process of the decomposition into exact and inexact 1-forms over successive submanifolds of an infinite-dimensional manifold.